\documentclass[12pt]{amsart}
\usepackage[utf8]{inputenc}
\usepackage{amsfonts}
\usepackage{amsmath}
\usepackage{amssymb}
\usepackage{amsthm}
\usepackage{xcolor}
\usepackage[margin=1.3in]{geometry}
\usepackage{graphicx}
\usepackage[11pt]{moresize}

\newtheorem{conjecture}{Conjecture}

\theoremstyle{remark}

\newtheorem*{remark*}{Remark}

\newcommand{\vs}{\vspace{3mm}}

\newcommand{\sub}{\subseteq}

\title{Experimental Evidence for Asymptotic Non-Optimality of Comb Adversary Strategy}
\author{Zachary Chase}
\address{Mathematical Institute, Andrew Wiles Building, Radcliffe Observatory Quarter, Woodstock Road, Oxford OX2 6GG, UK}
\email{zachary.chase@maths.ox.ac.uk}
\date{December 3rd, 2019}

\begin{document}

\begin{abstract}
For the problem of prediction with expert advice in the adversarial setting with finite stopping time, we give strong computer evidence that the comb strategy for $k=5$ experts is not asymptotically optimal, thereby giving strong evidence against a conjecture of Gravin, Peres, and Sivan. 
\end{abstract}

\maketitle

\section{Introduction}

Prediction with expert advice is a classic setup in learning theory. Fix positive integers $k \ge 2$ and $T \ge 1$. On each of $T$ consecutive days, an \textit{adversary} assigns gains --- $0$ or $1$ --- to each of $k$ experts, and a \textit{player} who does not know the assigned gains for that day must choose an expert. The player receives the gain assigned to the chosen expert and then sees the gains all the experts were assigned that day. After $T$ days, each expert has a total gain, namely, the sum of the gains assigned to her each day, and the player similarly has a total gain. The \textit{regret} that the player has after $T$ days is the difference between the biggest total gain that an expert has and the total gain of the player. For a more formal problem statement, see [4]. 

\vs

With the adversary wishing to maximize the regret of the player and the player wishing to minimize his regret, we seek to find the minimax value of the game. Allowing the adversary and player to randomize their strategies, one can see that von Neumann's minimax theorem applies and confirms that there is indeed a minimax value of the game. The regime we will discuss is $k$ fixed (and small) and $T \to \infty$. It is well-known that the minimax regret value is given as $c_k\sqrt{T}+o(\sqrt{T})$ for some constant $c_k$ depending only on $k$. The value $c_2$ was found in 1965 by Cover [3]. In [4], Gravin, Peres, and Sivan gave another derivation of $c_2$ by finding a connection to random walks. In [1], Abbasi-Yadkori, Bartlett, and Gabillon found $c_3$. 

\vs

Another setting to consider the prediction with expert advice problem in is the ``geometric horizon model", in which, instead of having a known number $T$ days (``finite horizon model"), there is a (stopping) parameter $\delta$ such that the game ends independently each day with probability $\delta$. This setting is easier to analyze and is believed to be equivalent to the finite horizon model (see section 7 of version 4 of the arXiv version of [4]). In this setting, the minimax regret value is given as $\widetilde{c_k}\frac{1}{\sqrt{2\delta}}+o(\frac{1}{\sqrt{2\delta}})$ as $\delta \to 0$ with $k$ fixed. In [4], Gravin et. al. obtained $\widetilde{c_k}$ for $k=2,3$, and in [2], Bayraktar, Ekren, and Zhang obtained $\widetilde{c_4}$ using PDE methods. 

\vs

Gravin et. al. found a convex polytope with a small number of vertices such that a (minimax) optimal adversary strategy lies in this convex polytope. Essentially, determining the optimal adversary strategy comes down to determining the best way to locally couple $k$ random walks of fixed drift to maximize how far the farthest walk is away from the origin. They defined the ``comb strategy" as follows. At any given day $t$, the experts are ranked $1,2,\dots,k$ based on their current total gain, with $1$ having the highest current total gain. The adversary, with probability $\frac{1}{2}$, assigns a gain of $1$ to all odd ranked experts and a gain of $0$ to all even ranked experts, and with probability $\frac{1}{2}$, assigns a gain of $1$ to all even ranked experts and a gain of $0$ to all odd ranked experts. Gravin et. al. made the following conjecture.

\begin{conjecture}
For any $k \ge 2$, the comb strategy is asymptotically optimal, i.e., the (expected) regret resulting from the adversary employing the comb strategy has growth rate $c_k\sqrt{T}+o(\sqrt{T})$ as $T \to \infty$. 
\end{conjecture}

That the comb strategy is asymptotically optimal was proven for $k=2$ in [4] and $k=3$ in [1] for the finite horizon model, and for $k=2,3$ in [4] and $k=4$ in [2] in the geometric horizon model. 

\vs

In thise short note, we give strong computer evidence that the comb strategy is not asymptotically optimal for $k=5$ in the finite horizon model. As mentioned previously, it is believed that this is then strong computer evidence that the comb strategy is not asymptotically optimal for $k=5$ in the geometric horizon model. 

\vs

We compare the comb strategy, denoted $[1,3,5]$, to the following strategy, denoted $[1,3]$. The adversary, with probability $\frac{1}{2}$, assigns a gain of $1$ to experts $1$ and $3$ and a gain of $0$ to experts $2,4,5$, and with probability $\frac{1}{2}$, assigns a gain of $1$ to experts $2,4,5$ and a gain of $0$ to experts $1,3$, where, to reiterate, the labels of the experts are their rankings for the current day, based on total gain. 

\section{Computer Evidence}

For values of $T$ up to $350$, we compute exactly the expected regret from following the strategies $[1,3]$ and $[1,3,5]$, by using a simple dynamic program. If the comb strategy, $[1,3,5]$ were asymptotically optimal, then it would not be the case that $\frac{R_{1,3}(T)^2-R_{1,3,5}(T)^2}{T}$ is positive and bounded away from $0$, where $R_{1,3}(T)$ denotes the expected regret of the adversary employing strategy $[1,3]$ on day $T$ (and $R_{1,3,5}(T)$ is similarly defined). However, the extreme constancy of $\frac{R_{1,3}(T)^2-R_{1,3,5}(T)^2}{T}$ above $0$ suggests that it is. See figure $1$ below. 

\vs

\hspace{20mm} \includegraphics[scale=.8]{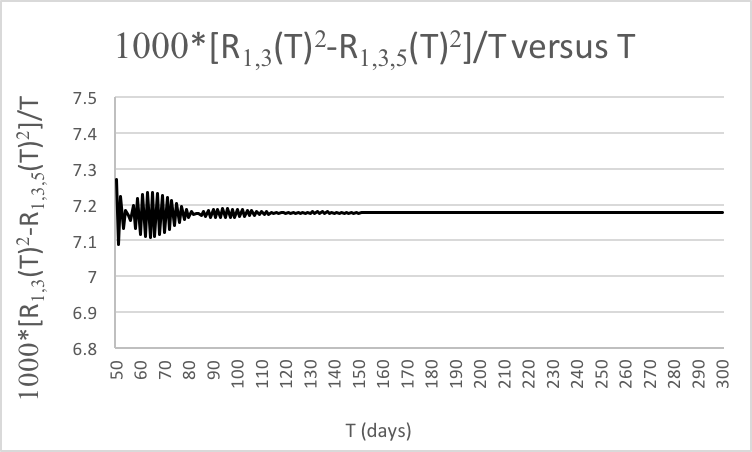}

\noindent {\ssmall \textit{Figure 1}: Difference in 1000 times the normalized regret squared between the comb strategy $[1,3,5]$ and the strategy $[1,3]$.}

\vs

It is straightforward to compute that for $T=5$, stategy $[1,3]$ is strictly better than strategy $[1,3,5]$. This is in contrast to the cases of $k=2,3,4$ in which doing comb appears by code to be always exactly optimal (for any $T$). It does appear from code that, for $k=5$, doing $[1,3]$ is always exactly optimal (for any $T$). However, it's not the case that, for $k=6$, there is one subset $A \sub \{1,2,3,4,5,6\}$ such that doing the strategy of giving a gain of $1$ to the experts ranked in $A$ and $0$ to the rest with probability $\frac{1}{2}$, and doing the opposite with probability $\frac{1}{2}$, is always exactly optimal. Indeed, one can run a dynamic program to see that, for $T=13$, if the strategies $[1,3,6]$ and $[1,4,6]$ are available on each day, then an expected regret of $9.14453125$ can be achieved, whereas the largest expected regret when just using one subset $A$ (such as $[1,3,6]$) that can be achieved is $9.143310546875$. This leads the author to believe that any asymptotically optimal strategy for $k=6$ experts will be quite complicated and in particular be heavily dependent on the particular current total gains, rather than just on their ordering.


\begin{thebibliography}{10}

\bibitem{c} Yasin Abbasi-Yadkori, Peter L. Bartlett, and Victor Gabillon, Near minimax optimal players for the finite-time 3-expert prediction problem. \textit{Advances in Neural Information Processing Systems 30, pages 3033–3042, Long Beach, California, 2017}.

\bibitem{d} Erhan Bayraktar, Ibrahim Ekren, and Yili Zhang, On the asymptotic optimality of the comb strategy for prediction with expert advice. 2019. \textit{URL https://arxiv.org/abs/1902.02368}.

\bibitem{a} Thomas M. Cover, Behavior of sequential predictors of binary sequences. \textit{Proceedings of the 4th Prague Conference on Information Theory, Statistical Decision Functions, Random Processes, pages 263–272, 1965}.

\bibitem{b} Nick Gravin, Yuval Peres, and Balasubramanian Sivan, Towards optimal algorithms for prediction with expert advice. \textit{Proceedings of the Twenty-Seventh Annual ACM-SIAM Symposium on Discrete Algorithms, pages 528–547, Arlington, Virginia, 2016}.

\end{thebibliography}
\end{document}